\documentstyle[epsfig,12pt,dina4p]{article}
\def\@oddhead{}\def\@evenhead{}
\def\@oddfoot{}
\def\@evenfoot{\@oddfoot}
\def\gev{\rm GeV}

\def\rap{\hat{y}}
\def\clb#1 {(#1 Coll.),}
\title{Photoproduction of $D^{*\pm}$ Mesons in $ep$ Collisions at HERA}
\author{H1 Collaboration}
\begin{document}

\begin{titlepage}
\begin{flushleft}
{\rm  DESY 96-055}\\  
{\rm  April 1996} \\ 
\end{flushleft}
\vspace*{4.cm}
\begin{center}
\begin{Large}
\boldmath
\bf{Photoproduction of $D^{*\pm}$ Mesons in Electron-Proton Collisions at HERA}
\unboldmath 

\vspace*{2.cm}
H1 Collaboration 
\end{Large}
\end{center}

\vspace*{3cm}
\begin{abstract}
At the electron-proton collider HERA 
the inclusive $D^{*\pm}$ meson photoproduction cross section 
has been measured with the H1 detector 
in two different, but
partly overlapping, kinematical regions.
For the first, where
$\langle W_{\gamma p}\rangle \approx 200$\ GeV and $Q^2 < 0.01\,\gev^2$,
the result is
$\sigma(\gamma p \rightarrow c \bar{c} X) 
= (13.2 \pm 2.2 ^{+2.1}_{-1.7}\, ^{+9.9}_{-4.8})\,\mu b$.
%
%
The second measurement for
$Q^2 < 4\,\gev^2$ yields
$\sigma(\gamma p \rightarrow c \bar{c} X) 
= ( 9.3 \pm 2.1  ^{+1.9}_{-1.8}\,  ^{+6.9}_{-3.2} )\,\mu b$
at $\langle W_{\gamma p}\rangle \approx 142$\,GeV and
$\sigma(\gamma p \rightarrow c \bar{c} X) 
= ( 20.6 \pm 5.5 ^{+4.3}_{-3.9}\, ^{+15.4}_{-7.2})\,\mu b$
at $\langle W_{\gamma p}\rangle \approx 230$\,GeV, respectively. 
The third error accounts for an additional uncertainty 
due to the proton and photon parton density parametrizations.
Differential cross sections 
are presented as a function of the $D^{*\pm}$ 
transverse momentum and rapidity.
The results compare reasonably well
with next-to-leading order QCD calculations.
Evidence for diffractive photoproduction of
charm quarks is presented.
\end{abstract}
\end{titlepage}

%
\noindent
\footnotesize
 S.~Aid$^{14}$,                   
 V.~Andreev$^{26}$,               
 B.~Andrieu$^{29}$,               
 R.-D.~Appuhn$^{12}$,             
 M.~Arpagaus$^{37}$,              
 A.~Babaev$^{25}$,                
 J.~B\"ahr$^{36}$,                
 J.~B\'an$^{18}$,                 
 Y.~Ban$^{28}$,                   
 P.~Baranov$^{26}$
% ...

\vspace*{0.5cm}
\scriptsize\noindent
% ...
$ ^{12}$ DESY, Hamburg, Germany$^a$ \\
% ...
$ ^{14}$ II. Institut f\"ur Experimentalphysik, Universit\"at Hamburg,
                                     Hamburg, Germany$^ a$  \\
% ...
$ ^{18}$ Institute of Experimental Physics, Slovak Academy of
                Sciences, Ko\v{s}ice, Slovak Republic$^ f$\\
% ...
$ ^{25}$ Institute for Theoretical and Experimental Physics,
                                                 Moscow, Russia \\
 $ ^{26}$ Lebedev Physical Institute, Moscow, Russia$^ f$ \\
% ...
$ ^{28}$ LAL, Universit\'{e} de Paris-Sud, IN2P3-CNRS,
                            Orsay, France\\
 $ ^{29}$ LPNHE, Ecole Polytechnique, IN2P3-CNRS,
                             Palaiseau, France \\
% ...
$ ^{36}$ DESY, Institut f\"ur Hochenergiephysik,
                              Zeuthen, Germany$^ a$\\
 $ ^{37}$ Institut f\"ur Teilchenphysik,
          ETH, Z\"urich, Switzerland$^ i$\\
% ...
$ ^a$ Supported by the Bundesministerium f\"ur Bildung, Wissenschaft,
        Forschung und Technologie, FRG,
        under contract numbers 6AC17P, 6AC47P, 6DO57I, 6HH17P, 6HH27I,
        6HD17I, 6HD27I, 6KI17P, 6MP17I, and 6WT87P \\
% ...
$ ^c$ Supported by FNRS-NFWO, IISN-IIKW \\
% ...
$ ^f$ Supported by the Deutsche Forschungsgemeinschaft\\
% ...
$ ^i$ Supported by the Swiss National Science Foundation\\
\normalsize
\newpage
\section{Introduction}

The study of heavy quark production in lepton-proton scattering 
provides an important testing ground for the standard model
\cite{ali}.
At the electron-proton collider HERA, 
heavy quarks are produced, 
according to QCD, by  direct and hadronic (resolved) photon processes.
The direct {\em photon gluon fusion} process 
$ \gamma g \rightarrow c\bar{c},\label{gagcc} $
where a photon emitted by the electron and a gluon
from the proton generate a $c\bar{c}$ pair is expected to 
dominate.
The major
contribution is due to the exchange of an almost real photon
{\em (photoproduction)}, 
where the negative squared four-momentum transfer carried by the photon 
is $Q^2 \approx 0$.
The scattered electron is either lost in the beampipe or
detected at small angles 
with respect to the electron beam direction. 
The fraction of $c\bar{c}$ events where the scattered electron is seen
in the main detector {\em (Deep Inelastic Scattering, DIS,
$Q^2 > 4\,\gev^2$)} is at least one order of magnitude 
smaller\, \cite{charm-dis}. 
Measurements at HERA can be considered as a 
continuation of fixed-target photoproduction
experiments \cite{fixed-target},
but at about one order of magnitude higher
centre-of-mass (CM) energies,
$W_{\gamma p} \sim {\cal O} (200)\,\gev$.

In the {\em Weizs\"acker-Williams Approximation} (WWA) \cite{wwa},
the electroproduction cross section $\sigma_{ep}$
is expressed as a convolution 
of the flux of photons emitted by the electron, $f_{\gamma/e}$,
with the photoproduction cross section 
\begin{equation}
 \sigma_{ep} = \sigma (e p \rightarrow e c \bar{c} X) = \int dy \ f_{\gamma/e}(y)
   \cdot \sigma (\gamma p \rightarrow c \bar{c} X), 
\label{wwa}
\end{equation}
where $y$ 
represents the fraction of the electron energy
transferred to the photon
in the proton rest frame.
For the {\em direct} photoproduction process, the 
cross section $\sigma_{\gamma p}$, in turn, 
is assumed in leading order to factorize into the
photon gluon fusion cross section and the gluon density in
the proton
\begin{equation}
 \sigma_{\gamma p} = \sigma (\gamma p \rightarrow c \bar{c} X) = 
   \int dx_g \ x_g g(x_g,\mu^2) \cdot \sigma(\gamma g \rightarrow c \bar{c}).
\label{gammap}
\end{equation}
Here $x_g$ denotes the momentum fraction of the proton carried by the
gluon and $\mu$ the factorization scale.
Estimates of the cross sections depend on
the behaviour of the 
gluon density distribution $g$ of the proton at small $x_g$,
on the QCD renormalization scale, on
the factorization scale, and on the heavy quark 
mass $m_c$~\cite{frixione-1}.

Next-to-leading order (NLO) corrections
to the parton cross section\,\cite{frixione-1,ellis} 
are found to be substantial,
but are reduced by experimental selection criteria, 
which limit the acceptance to the central rapidity
range in the $\gamma p$ CM system
at large transverse momenta~\cite{frixione-1}.

Charm photoproduction can also proceed via the
hadronic component of the photon {\em (resolved photon processes),}
where a parton inside the photon scatters off a
parton inside the proton, {\it e.g.}
$ g g \rightarrow c\bar{c}.\label{ggcc} $
This process known to dominate
light quark production                  
is  expected to contribute much less
to the production of charmed quark pairs\,\cite{frixione-1}. 
However, the production cross section still depends strongly
on the parton density function of the photon \cite{frixione-2}.
Other mechanisms, as for example the production of 
charm in the
fragmentation process, which is suppressed by the mass of the charm
quark, or
the production from the intrinsic charm content
of the nucleon \cite{schuler}, are believed to be small.
These processes, as well as any possible intrinsic charm component
of the photon,
are neglected in the present analysis.

Heavy quark production offers 
in principle the possibility of probing
the gluon distribution in the proton and the photon either 
indirectly, by measurement of the total
photoproduction cross section or of differential distributions,
or directly, by the explicit reconstruction of $x_g$. 
Measurements of the first kind are described here.
A similar measurement of $\sigma_{\gamma p}$ 
has also been reported by the 
ZEUS Collaboration \cite{dstar-zeus}.

The analysis makes use  of the $D^*$-tagging~\cite{feldmann},
{\it i.e.} of the tight kinematical conditions in the 
decay\footnote{Henceforth, 
charge conjugate states are always implicitly included.}
$ D^{*+} \rightarrow D^{0} \pi^{+}$, where the
$D^{0}$ mesons are reconstructed in the decay channel
$D^{0} \rightarrow K^{-}\pi^{+}$.
A better resolution is achieved in the distribution of the
mass difference
\begin{equation}
 \Delta M = M(D^0 \pi^+) - M(D^0)\label{deltam}
\end{equation}
than in the $D^{*+}$ mass distribution itself, whose
width is dominated by the momentum resolution of the detector.

The contribution of $D^{*+}$ mesons, originating from decays
of $b$ flavoured hadrons is neglected, 
because of the expected small $b$ production cross section 
at HERA ($\cal O$(5\,nb)\,\cite{eichler}).
$D^{*+}$ mesons from decays of higher-mass charm states
({\em e.g.} $D^{**})$ are not separated. 

Recently, much interest has been focused on a subclass of 
electroproduction events in which there is no hadronic
energy flow in an
interval of pseudorapidity, $\eta = - \ln \tan(\theta/2)$,
adjacent to the proton beam direction.
These diffractive processes are interpreted as
an exchange of a colour-less object with the quantum numbers
of the vacuum. 
The study of charm 
production in these processes is expected to provide information
on the partonic structure of diffractive exchange.

\section{Analysis}

The present analysis is based on data
collected with the H1 detector
during the 1994 running period of the HERA collider, where 
27.5\,\gev\ positrons
collided with 820\,\gev\ protons,
at a CM energy of 300\,\gev.
A detailed description of the detector and its
trigger capabilities can be found elsewhere \cite{detector}. 

\subsection{Detector Description}

Charged particles are 
measured by two cylindrical jet drift chambers \cite{cjc,cjc-res}, 
mounted concentrically around the
beamline inside a homogeneous magnetic field of 1.15 Tesla,
yielding particle charge and momentum from the track curvature
in the polar angular range\footnote{H1 is using a 
right-handed coordinate system with the $z$ axis pointing in
the direction of the proton beam (forward), the $x$ axis 
pointing towards the centre of the storage ring.
The direction of the incoming positron
beam is termed backward. The polar angle $\theta$ is measured
with respect to the proton direction.}
of 20$^{\circ}$ to 160$^{\circ}$.
%
Two double layers of cylindrical multiwire proportional 
chambers (MWPC) \cite{mwpc} with pad readout
for triggering purposes are positioned inside and 
in between the two drift chambers, respectively.
The tracking detector is surrounded by a fine grained liquid
argon calorimeter \cite{calo}, consisting of an electromagnetic
section with lead absorbers and a hadronic section with
steel absorbers.
It covers polar angles between 4$^{\circ}$ and 155$^{\circ}$.
The luminosity is determined from the rate of {\em Bethe-Heitler}
$e p \rightarrow e p \gamma$ bremsstrahlungs events.
The luminosity system
consists of an electron detector and a photon detector,
located 33 m and 103 m from the interaction point
in the positron beam direction,
respectively.
The electron detector is used to 
tag photoproduction
events by detecting positrons scattered at small angles.
A time-of-flight system (TOF) is located in the backward
direction at $z \approx -2$\,m. 


\subsection{Data selection and $D^{*+}$ reconstruction}

The analysis is carried out independently for the case 
where the scattered positron
is detected in the 
electron tagger
and for the case where it is not required to be seen.
Henceforth, the respective data samples will be referred to 
as {\it tagged} and {\it untagged} sample. 
They correspond to integrated luminosities of 
$(2.77\,\pm\,0.04)\,{\rm pb}^{-1}$
and $(1.29\,\pm\,0.02)\,{\rm pb}^{-1}$,
respectively. 
About 20\,\% of the reconstructed $D^*$ candidates in the
tagged sample are also present in the untagged sample.
 
Proton beam induced background is reduced by 
requiring the event vertex to lie within
$\pm 40$\,cm of the nominal interaction point
along the beam direction.
A further
reduction is achieved by excluding events with energy flow only
into the forward region of the detector.

%
%
For each event all possible $M(K^- \pi^+)$ 
mass combinations are calculated with tracks of 
transverse momenta $p_t > 0.5\,\gev/c$. 
No particle identification is applied; 
each particle is assumed to be a
kaon or a pion in turn. 
Pairs with an invariant mass within
$\pm 80$\,MeV/c$^2$ of the nominal $D^0$ mass of 
$1.865\,\gev/c^2$ are 
combined with an additional track with $p_t > 0.15\,\gev/c$
and a charge opposite to that of the kaon candidate.

\begin{figure}[p] \centering
\epsfig{file=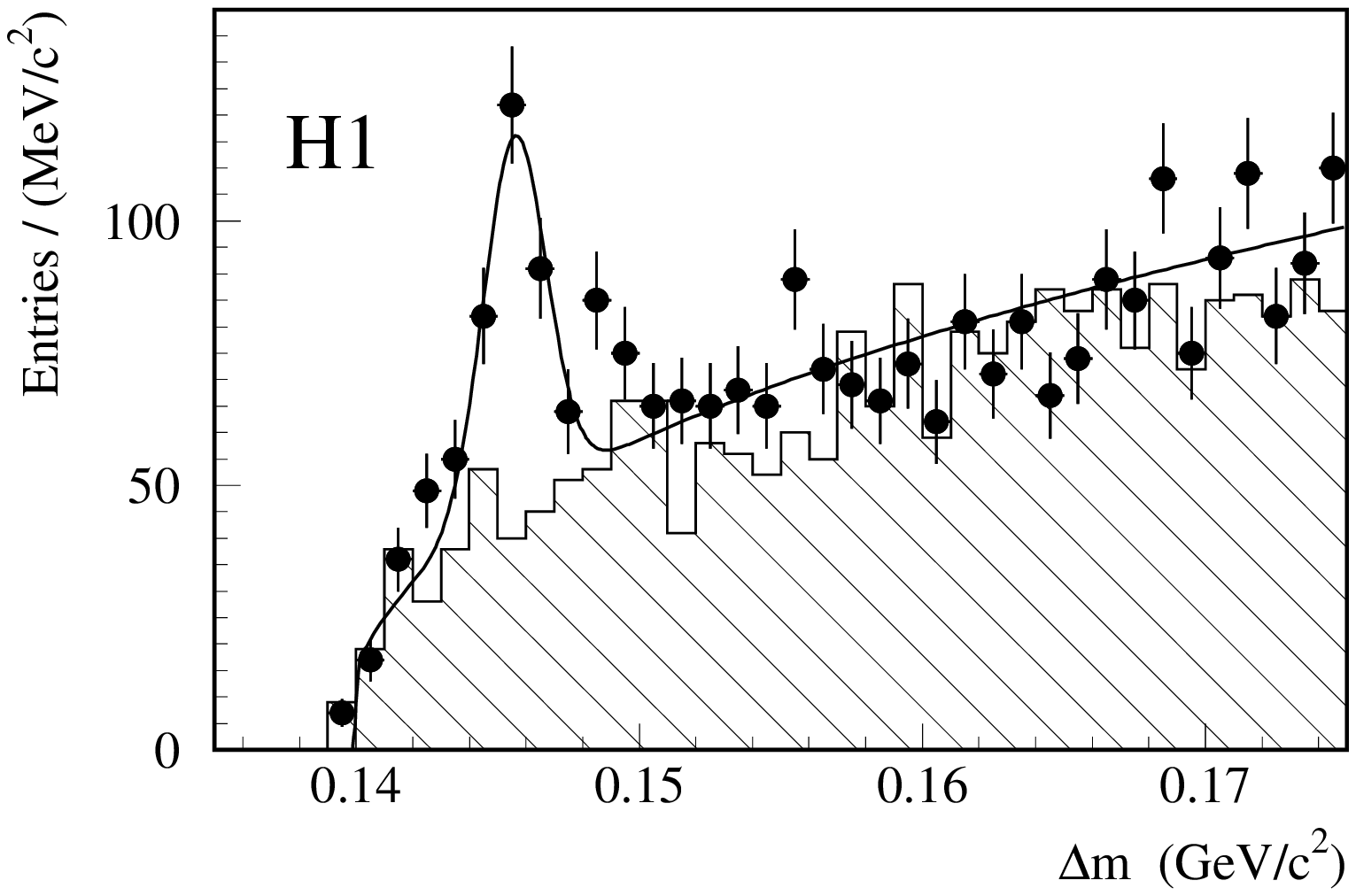,width=18cm}
\caption[Mass difference distribution]%
        {Distribution of the mass difference 
         $\Delta M = M(K^-\pi^+\pi^+) - M(K^-\pi^+)$
         for the combined tagged and untagged sample.
         The solid dots represent the data, the hatched
         histogram indicates the background as obtained from
         wrong charge combinations. The solid line is a fit of
         a Gaussian 
         plus a term for the background, as described in the text,
         with fixed width and position of the signal.}
\label{DMfigure}
\end{figure}

Figure\,\ref{DMfigure} shows the distribution of
the mass difference (\ref{deltam})
for $D^{*+}$-candidates 
with $p_t(D^{*+}) > 2.5\,\gev/c$ and a rapidity
$-1.5 < \rap(D^{*+}) =  - \frac{1}{2} \ln{\frac{E - p_z}{E + p_z}} < 1$ 
for the tagged and untagged samples combined. 
$D^{*+}$ production is found as a distinct enhancement,
containing about 190 combinations in a $\pm 2.5\,$MeV/c$^2$ window 
around the expected mass difference of $145.4$\,MeV/c$^2$.
No enhancement is observed if the mass difference for the
wrong charge combinations 
$ M(K^- \pi^- \pi^+) - M(K^- \pi^-)$
is used instead, as shown by the shaded histogram 
in figure\,\ref{DMfigure}.
The number of $D^*$ candidates is obtained 
from a simultaneous fit to signal and background 
events in the right-sign (RS) and wrong-sign (WS) distributions of
the $\Delta M$ spectra.
The signal is described by a Gaussian and the background
shape is parametrized by a function of the form
$a_i \cdot ( \Delta M - m_{\pi})^b $, (i=RS, WS).
The position and the width of the signal are determined from
a fit to a larger data sample using additional trigger conditions,
and then kept fixed at those values
($\Delta M_0 = 145.4$\,MeV/c$^2$,
$\sigma(\Delta M) = 1.11$\,MeV/c$^2$)
for all subsequent fits.
Uncertainties from variations of the fit procedure are 
accounted for in the systematic error. 

A Monte Carlo simulation  
is used to determine the efficiency for the 
reconstruction, the
selection cuts, and the acceptance of the detector.
Hard scattering events for direct and resolved
photoproduction are generated in leading order with the
PYTHIA 5.7 program~\cite{pythia}.
The generated events are 
fed into the H1 detector simulation program,
and are subjected to the same reconstruction and analysis
chain as the real data.

%


The tracking efficiencies have been examined in detail
using data.
The single track reconstruction efficiency
$\epsilon_{track}$               
is obtained by scanning tracks of high $p_t$ cosmic muons,
where the measured $p_t$ of the
incident track segment is compared 
with that of the outgoing track segment.
The $p_t$-dependence of $\epsilon_{track}$ is determined directly 
from the data by a novel 
method \cite{erdmann} based on the decay property 
of the pseudoscalar $K^{0}_{s}$ meson, decaying
isotropically  in its rest frame. 
The efficiency is
found to rise from 0 to the maximum value
between $p_t = 90$\,MeV/c and $p_t = 120$\,MeV/c, 
and to remain constant beyond that.
The precision measured in these studies
is quoted as the systematic error.
For single tracks the uncertainty found is
$\pm 2\% $ for the track reconstruction
and $^{+0}_{-3} \% $ for associating the track
to the primary vertex.
Combining them for the three tracks yields a
systematic error of $^{+6}_{-11} \%$ on the 
tracking efficiency.
%
 
\subsection{Analysis of electron tagged data}

Tagged events are required to have a positron candidate with
energy  $E_{e^{\prime}} > 4\,\gev$ in the electron tagger
and to have less than
$2\,\gev$ energy deposited in the photon detector.
In addition, 
at least one charged track candidate has to be detected by means
of a                 
MWPC trigger \cite{mwpc,ray-trig} and
a drift chamber track trigger \cite{dcrfi-trig}, 
thus ensuring activity in the central detector.
The trigger efficiency is determined from the data itself, using
independent triggers.
The analysis is restricted to the kinematical region 
$0.28 < y = 1 - E_{e^{\prime}}/E_e < 0.65$ and 
$Q^2 < 10^{-2}\,\gev^2$, where the
acceptance of the electron tagger is above $20\,\%$ with
an average value of about $60\,\%$. 
As a consequence, the $\gamma p$ CM energy range is
limited to 
159\,\gev $ < W_{\gamma p} < 242$\,\gev,
with a  mean of $W_{\gamma p} \approx 200\,\gev$ and an average
$\langle Q^{2}\rangle \approx 10^{-3}\,\gev^2$. 
The efficiency excluding the $y$ dependent electron tagger
acceptance is found to be 
$(48 \pm 4)\% $ and $(58 \pm 5)\% $ for direct and
resolved processes, respectively.
%

\subsection{Analysis of untagged data}
The untagged sample covers the kinematical region 
$0.1 < y < 0.8$ and $Q^2 < 4\,\gev^2$ at an average
$\langle Q^{2}\rangle \approx 0.2\,\gev^2$.
Contributions from DIS with 
$Q^2 > 4\,\gev^2$ are rejected by requiring that {\em no}
scattered positron candidate with $E_{e'} > 10$ GeV be
measured in the main detector.
The remaining background from DIS events is suppressed by
requiring $y < 0.8$ and is estimated
to be less than 1\,\%. 
Here $y$ is calculated from all measured final state particles
using the Jacquet-Blondel method~\cite{jb}.

%
The events are triggered by a combination of signals
from the central and rear parts of the detector.
At least one MWPC track candidate is 
required to point backwards into the region 
$110^{\small o}< \theta <155^{\small o}$.
The backward TOF system must
positively identify
the event as a genuine $ep$ collision 
by registration of a particle 
within the proper interaction time window
and within its angular acceptance of approximately
$160^{\small o}< \theta <177^{\small o}$.
The trigger efficiency for the central 
MWPC and drift chamber trigger components 
is determined by simulation 
to be $(84\pm 4)\%$.
For the backward part it is obtained from data, 
imposing the same selection criteria but using independent 
triggers based on local coincidences 
of MWPC tracks and low threshold ($>1.5\,\gev$) signals 
in the liquid-argon (LAr) calorimeter~\cite{bigray}.
Sufficient statistical precision to determine the efficiency 
in bins of $p_{t}$ and $\rap$ is achieved  
by including the sideband region of the mass difference signal, 
$0.15\,\gev/c^2 < \Delta M < 0.18\,\gev/c^2$, and 
the wrong sign combinations. 
To account for the different event topology of the 
combinatorial background
the efficiency is determined and parametrized as 
a function of $y$ 
and then folded with the $y$ spectrum of simulated $D^{*}$ events.
This yields efficiencies of the backward trigger component  
of $(28 \pm 4)\%$ for direct and $(35 \pm 5)\%$ for resolved 
production processes, respectively. 

%
\section{Results}

The visible production cross section
in $ep$ collisions is calculated from the
observed number of $D^{*+}$ mesons, $N$,
in the kinematical ranges of $p_t(D^*)> 2.5\,\gev/c$ and
rapidity $-1.5 < \rap(D^*) < 1$
according to the formula
\begin{equation}
\sigma_{D^{*\pm}} = 
\sigma(ep \rightarrow D^{*\pm} X) = 
\sigma(ep \rightarrow D^{*+} X) + \sigma(ep \rightarrow D^{*-} X) =
\frac{N}
  { L \cdot B \cdot \epsilon },
\label{sigman}
\end{equation}
where  $L$ denotes the integrated luminosity, $\epsilon$ the total
efficiency, and
$B = B(D^{*+} \rightarrow D^0 \pi^+) \cdot B(D^0 \rightarrow K^- \pi^+) =
0.0273 \pm 0.0011$ \cite{PDG} is 
the combined branching fractions of $D^{*+}$ and $D^0$ mesons.
For the analysis of the tagged sample, 
the acceptance of the electron tagger and its
trigger efficiency are accounted for on an event-by-event basis.
For the relative ratio of direct to resolved
photoproduction processes
the values predicted by the NLO QCD calculation are used
({\it i.e.} 79:21 for the full $(\rap, p_{t})$ range
or 93:7 for the visible kinematical range).
The charm quark mass is assumed to be $m_c =1.5$ GeV/c$^2$, 
the ratios of the factorization scale for the photon, the
factorization scale for the proton, and the renormalization scale 
are taken to be $2 m_c, 2 m_c$, and $m_c$, 
as recommended by the authors\,\cite{frixione-1}.

\subsection{Visible cross section $\sigma_{e p}$} 

%
In the tagged sample
the fitted number of $D^{*}$ mesons, $119 \pm 16$,
is corrected for the electron tagger
acceptance~\cite{sig-tot},
yielding $N = 197 \pm 28$.
For the kinematical region 
$p_t(D^*)> 2.5\,\gev/c^2$,
$-1.5 < \rap(D^*) < 1$, $Q^2 < 0.01$ GeV$^2$,
and $159 < W_{\gamma p} < 242\,\gev$ 
the {\em visible} $ep$ production cross section is
determined to be
\begin{equation}
 \sigma(e p \rightarrow D^{*\pm} X) = 
(4.90 \pm 0.70 ^{+0.74}_{-0.59} )\,{\rm nb} \hspace{2.5cm} {\rm (tagged)}, 
\label{sigep}
\end{equation}
where the errors refer to the statistical 
and the experimental systematic error (see below).
%

%
In the case of untagged events, 
the fitted number of $D^{*}$ mesons is
$97 \pm 15$ events and
the average total efficiency for $95 < W_{\gamma p} < 268\,\gev$ and
$Q^2 < 4\,\gev^2$ is found to be $0.14 \pm 0.03$.
The $ep$ cross section in the kinematical region
$p_t(D^*)> 2.5\,\gev/c^2$,
$-1.5 < \rap(D^*) < 1$ is thus measured as
\begin{equation}
 \sigma(e p \rightarrow D^{*\pm} X) 
= (20.2 \pm 3.3 ^{+4.0}_{-3.6})\, {\rm nb} \hspace{3.0cm} {\rm (untagged)}.
\label{sigep-untag}
\end{equation}

The visible cross section 
is almost insensitive to both 
the choice of parton density parametrizations and
to the mixture
of direct and resolved photoproduction processes,
because the efficiencies are very similar 
and there is no  acceptance correction involved.

%
The experimental systematic uncertainties 
are listed in table~\ref{tab-syserr}.
In the analysis of tagged data, the largest contribution 
($11\,\%$) is due to the uncertainty
in the track reconstruction,
whereas in the untagged case 
the largest error arises from 
a $14\,\%$ uncertainty in the determination of the trigger efficiency. 
Adding the various uncertainties in quadrature results in a total
experimental systematic error for the tagged sample of 
$^{+15}_{-12} \%$ for the inclusive $D^*$ cross
section and $^{+16}_{-13}\,\%$ for the charm cross section (see 
below), respectively. 
For the untagged sample the corresponding 
uncertainties are
$^{+20}_{-18}\,\%$ and $^{+21}_{-19}\,\%$.

\begin{table}
\begin{center}
\begin{tabular}{|l|r|r|}
\hline
        & \hspace*{0.3cm} Tagged  & \hspace*{0.3cm} Untagged  \\
\hline
Track trigger    & 5\ $\%$              & 5\ $\%$\\
Electron tagger acceptance       & 5 $\%$               & ---  \\
Backward trigger        & ---              & 14 $\%$ \\
Track reconstruction    & $^{+11}_{-6} \%$  &   $ ^{+11}_{-6} \%$ \\
Signal extraction/background subtraction & 6\ $\%$              & 6\ \% \\
Luminosity               & 1.5\ $\%$            & 1.5\ $\%$\\
$D^*, D^0$ branching ratios     & 4\ \% & 4\ \%\\
$c \rightarrow D^*$ branching fraction & 7\ \% & 7\ \%\\
\hline
  Total experimental uncertainty  & $ ^{+16}_{-13} \%$  &  $ ^{+21}_{-19}\%$   \\
\hline 
\end{tabular}
\caption[Summary experimental systematic uncertainties]
{Experimental systematic uncertainties.}
\label{tab-syserr}
\end{center}
\end{table}

Predictions by the NLO QCD calculation for
the visible cross  section $\sigma_{D^{*\pm}}$
in the tagged case assuming the
following pairs of proton and photon parton densities of 
(GRV HO\cite{grv-ho} + GRV-G HO\cite{grv-ho}),
(MRSA'\cite{mrsap} + GRV-G HO), 
(MRSD0'\cite{mrsdprime} + GRV-G HO) and 
(MRSA' + LAC1\cite{lac1})
yield values of 3.2, 2.8, 2.4 and 
2.8\,nb respectively.
With the present accuracy
a clear distinction is not possible, albeit
a slightly better agreement is reached for parton densities
with a rising gluon density distribution at low $x_g$.
A similar conclusion results from the analysis of the untagged data.
This is in agreement with measurements of the 
gluon density by other methods~\cite{gluon},
{\em e.g.}\ from scaling violations
of $F_2$ in DIS.

%
\subsection{Total cross sections $\sigma_{ep}$ and $\sigma_{\gamma p}$}

%

The visible cross sections (within a limited
$(\rap, p_t)$ phase space) have to be
extrapolated to the full $(\rap, p_t)$ phase space
to obtain the total cross sections.

The individual acceptances for the
direct and resolved processes as well as
their relative strength depend on the
choice of the parton densities, and therefore
so also does the extrapolation performed by simulation.
This is illustrated in table\,\ref{tab-accep},
which lists values for the  acceptance
calculated for various parton densities of the proton and photon 
and for different charm quark masses, 
for the kinematical region of the tagged sample. 
The numbers for the untagged case are similar. 
The acceptance is defined as the fraction of 
$D^*$ mesons within the
quoted $\rap$ and $p_t$ ranges with respect to
the total number of produced~$D^*$.

The derivation of the total cross sections is
based on simulations using
a charm quark mass of 1.5\,\gev/c$^2$
and assuming the GRV LO \cite{grv-lo}
parametrizations for both the proton and photon
parton densities,
which are in good agreement with measured 
parton densities.
This leads to a  
charm $ep$ production cross section of
$\sigma_{e p} =
(941 \pm 160 ^{+142}_{-120})\,$ nb
at $\sqrt{s} = 300\,\gev$ and $Q^2 < 0.01\,\gev^2$
for the full $y$-range. 
The effect of hadronization is included using the 
fragmentation fraction $B_{c \rightarrow D^{*+}} = 0.260 \pm 0.021$ 
\cite{OPAL-cd}.
The $ep$-cross section is converted into a $\gamma p$ cross 
section  using equation~(\ref{wwa}), 
which yields for $Q^2 < 0.01\,\gev^2$
\begin{equation}
 \sigma(\gamma p \rightarrow c {\bar c} X) 
= ( 13.2 \pm 2.2 ^{+2.1}_{-1.7}\, ^{+9.9}_{-4.8})\,\mu {\rm b}  \qquad  \qquad
{\rm at} \quad \langle W_{\gamma p}\rangle \approx 200\,\gev.
\label{siggp-eq1}
\end{equation}
For the untagged case the result over the range of
$95\,\gev < W_{\gamma p} < 268\,\gev$ becomes
\begin{equation}
 \sigma(\gamma p \rightarrow c {\bar c} X)
= ( 12.6 \pm 2.1 ^{+2.6}_{-2.4}\, ^{+9.4}_{-4.4} )\,\mu {\rm b}  \qquad \qquad
 {\rm at} \quad \langle W_{\gamma p}\rangle \approx 180\,\gev.
\label{siggp-eq4}
\end{equation}

The third error indicates the additional extrapolation uncertainty
as discussed below.
The larger available kinematic range in 
$ W_{\gamma p}$ allows a division into the two regions
$95\,\gev < W_{\gamma p} < 190\,\gev$ and
$190\,\gev < W_{\gamma p} < 268\,\gev$, thus providing
information about the energy dependence of the cross section.


\begin{table}
\begin{center}
\begin{tabular}{|c|c|c||c|}
\hline
Proton         &     Photon     & $m_{c}$   &            \\
parton density & parton density & [GeV/c$^2$] & Acceptance\\
\hline
\hline
GRV LO  \cite{grv-lo} & --- & $1.2$ & $4.8  \,\,\% $ \\
GRV LO  \cite{grv-lo} & --- & $1.5$ & $6.3 \,\,\% $\\
GRV LO \cite{grv-lo} & --- & $1.8$ & $10.8 \,\,\% $\\
\hline
\hline
MRSG \cite{mrsa} & ---  & $1.5$  & $6.7 \,\,\% $\\
MRSA' \cite{mrsap} & ---   & $1.5$ & $10.4 \,\,\% $\\
\hline
\hline
GRV LO \cite{grv-lo}  & GRV LO \cite{grv-lo} & $1.5$  & $2.1 \,\,\% $\\
GRV LO \cite{grv-lo}  & LAC1 \cite{lac1}& $1.5$ & $0.7 \,\,\% $\\
\hline
\hline
\end{tabular}
\caption{Acceptance for different parton density
parametrizations for the direct or resolved contributions,
respectively, and for different charm quark masses, 
as used in the extrapolation from the visible to
the total cross section.}
\label{tab-accep}
\end{center}
\end{table}

\begin{figure}[p] \centering 
\epsfig{file=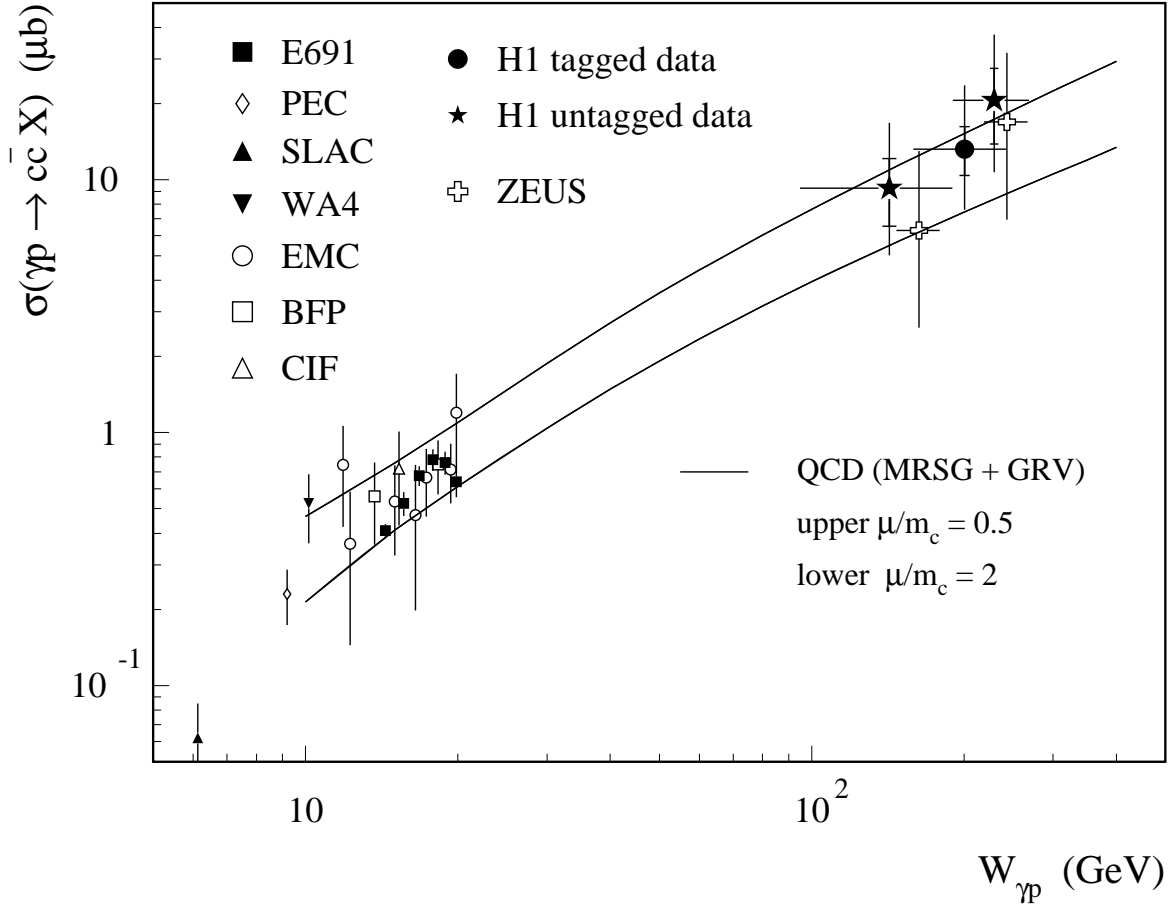,width=17cm,clip=}
\caption[$W_{\gamma p}$ dependence of cross section ]%
            {Total charm photoproduction  cross section 
             as a function of $W_{\gamma p}$.
             The solid dots and  stars
             represent the present analyses with
             statistical and systematic errors added 
             in quadrature (inner error bars). 
             The outer error bars  indicate the total
             error if in addition the uncertainty due to 
             the choice of parton density parametrizations
             is added in quadrature.
             The crosses refer to the results of the ZEUS collaboration, 
             the other symbols indicate earlier measurements at 
             fixed-target experiments. 
             The solid lines represent the prediction
             of a NLO QCD calculation using the MRSG and GRV-G HO
             parametrizations of the proton and photon parton densities,
             respectively. The upper and lower lines delimit 
             the range of values expected from
              varying the renormalization scale
              within $0.5 < \mu/m_c < 2.$}
\label{wgpfigure}
\end{figure}

The results are summarized in table~\ref{tab-summary}
for both analyses and
compared in figure\,\ref{wgpfigure}      
with measurements by the
ZEUS collaboration (at similar $W_{\gamma p})$ \cite{dstar-zeus}, 
and previous
fixed-target experiments at lower energies~\cite{fixed-target}.
The inner error bars represent the statistical and experimental
systematic errors added in quadrature. 
The outer set of error bars indicate the total error after 
adding in quadrature the extrapolation 
uncertainty discussed below.
The cross section is rising by almost one order of magnitude
as compared to the low energy measurements.

Overlaid in figure\,\ref{wgpfigure} are predictions
by the NLO QCD calculation~\cite{frixione-1} 
with parton density parametrizations MRSG~\cite{mrsa}
for the proton and GRV-G HO~\cite{grv-ho} for the photon. 
The upper and the lower solid lines delimit the range of 
values expected due to a variation of the 
renormalization scale within $m_c/2 < \mu < 2 m_c$.

%
%
\begin{table}
\begin{center}
\begin{tabular}{|l|r|r|r|}
\hline
 Quantity                & Tagged        & Untagged   & Untagged \\
\hline
Range in $W_{\gamma p}$  (GeV)   & 159 - 242   &  95 - 190  & 190 - 268  \\
Range in $Q^2$  (GeV$^2$)   & $ < 10^{-2} $ &  $< 4$   &   $< 4$    \\
$D^*$ candidates          & 119 $\pm 16$ &  $ 46\pm 9$ &  $51 \pm 12$ \\
$ \langle\epsilon_{tot}\rangle (\%)$ (direct)  & $ 29 \pm 2$ & $11 \pm 1$  & $17 \pm 2$ \\
Photon flux              &  0.0141     & 0.0486  & 0.0155  \\
\hline
$\sigma(ep \rightarrow D^* X)$ [$y, Q^2, p_t, \rap$] (nb)
                 & $4.90 $   & 11.0  &  8.5  \\
\quad Errors     & $\pm 0.70 ^{+0.74}_{-0.59}$   
                 & $\pm 2.4 ^{+2.2}_{-2.0}$
                 & $ \pm 2.2 ^{+1.7}_{-1.5}$  \\
\hline 
$\sigma(\gamma p \rightarrow c{\bar c} X)$ [$y, Q^2$] ($\mu b$)
     &   13.2  & 9.3 & 20.6 \\
\quad Errors  & $ \pm 2.2 ^{+2.1}_{-1.7}\,  ^{+9.9}_{-4.8} $  
              & $ \pm 2.1 ^{+1.9}_{-1.8}\,  ^{+6.9}_{-3.2} $ 
              & $ \pm 5.5 ^{+4.3}_{-3.9}\,  ^{+15.4}_{-7.2}$  \\
\hline 
\end{tabular}
\caption[Summary cross section results]
{Cross section results for tagged and untagged data
samples. Errors shown are statistical, experimental systematic, and
for $\sigma(\gamma p \rightarrow c{\bar c} X)$ also
uncertainties due to the dependence on the 
parton density parametrizations.}
\label{tab-summary}
\end{center}
\end{table}

Calculating $\sigma_{\gamma p}$ with other combinations
of parton density parametrizations
increases the measured $\sigma_{\gamma p}$
by up to 75\% (in the case of MRSG and LAC1), or decreases
$\sigma_{\gamma p}$ by up to 35\% (in the case of MRSA' and GRV-G HO).
This variation reflects
the uncertainties due to the choice of parton densities, 
and is quoted separately as a third error in the
$\sigma_{\gamma p}$ cross sections. 
The extrapolation uncertainty due to fragmentation models
($ < 30 \%$, estimated by a comparison with a
cluster fragmentation as 
implemented in the Herwig program \cite{herwig})
and due to the choice of the charm quark mass (see table 2)
is not included in the error.

If the extrapolation is based on the
proton parton density parametrizations
MRSG, MRSA'
 or  MRSD0' \cite{mrsdprime}, the value of 
$\sigma_{\gamma p}$ obtained 
(for the tagged sample at $W_{\gamma p}=200$ GeV) becomes
$(12.2 \pm 2.0 ^{+2.0}_{-1.6})\,\mu$b,
$(8.6 \pm 1.4 ^{+1.4}_{-1.1})\,\mu$b or
$(7.4 \pm 1.2 ^{+1.2}_{-1.0})\,\mu b$,  which are
to be compared 
with the QCD predictions of 
9.8\,$\mu$b, 6.0\,$\mu$b or 3.9\,$\mu$b, respectively.
Measurement and prediction change in the same manner.
Hence a total cross section measurement 
can presently not distinguish 
between the different gluon densities.

\subsection{Differential distributions}
\begin{figure}[p] \centering
\epsfig{file=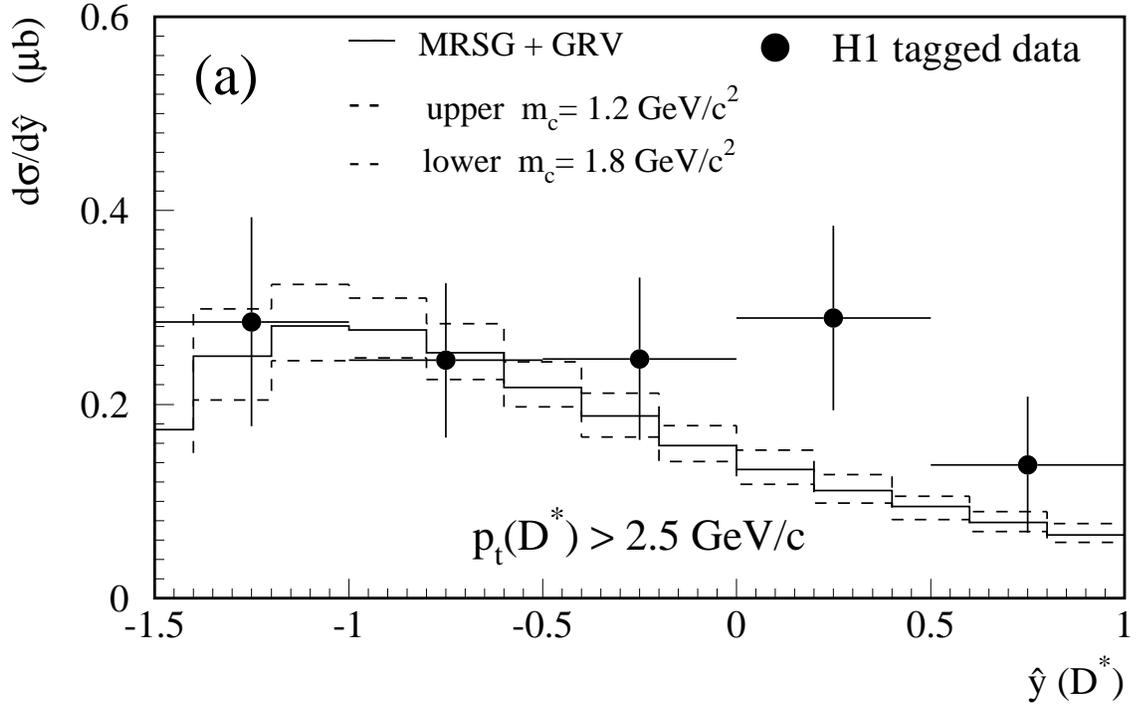,width=17.cm}
\epsfig{file=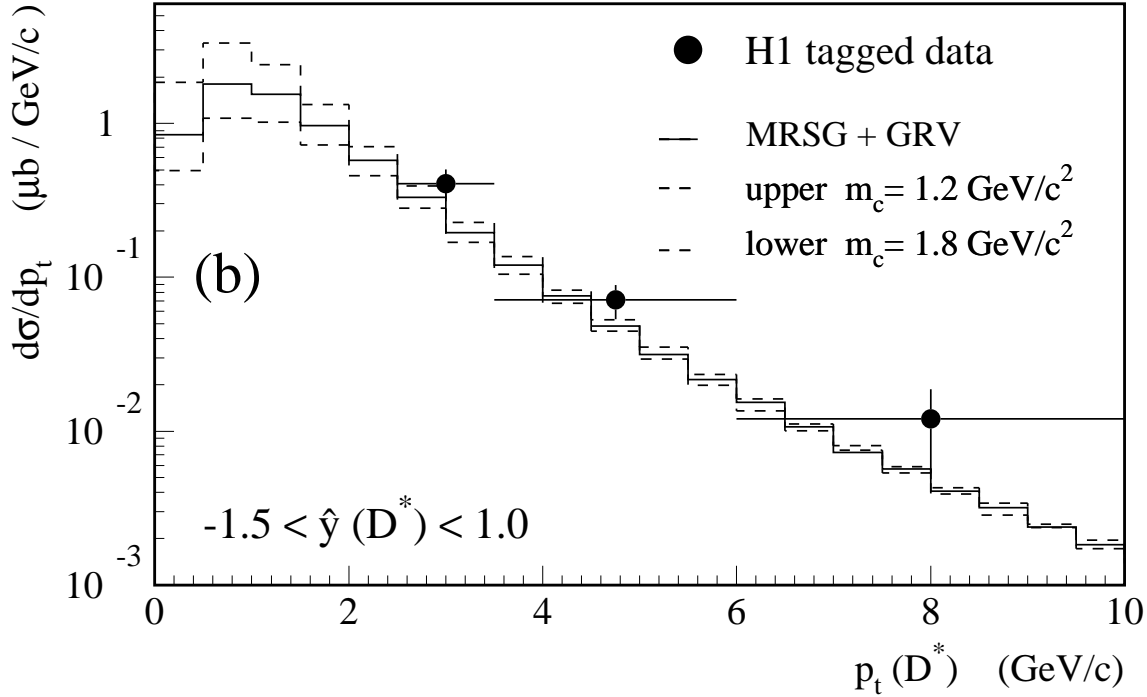,width=17.cm}
\caption[Differential eta]%
        {Differential cross sections for the tagged sample 
         (solid dots).
         (a) $1/(2 B_{c \rightarrow D^{*+}}) \cdot d\sigma 
              (\gamma p \rightarrow D^{*\pm} X)/ d\rap$
         for events with $p_t(D^*) > 2.5$ GeV/c and 
         (b) $1/(2 B_{c \rightarrow D^{*+}}) \cdot d\sigma 
               (\gamma p \rightarrow D^{*\pm} X)/ dp_t$
         for events with $-1.5 < \rap (D^*) < 1.$
         The solid histogram shows the NLO QCD prediction, using
         the MRSG proton parton density parametrization
         with a charm quark mass of 1.5 GeV/c$^2$.
         The upper (lower) dashed histogram
         indicates the effect of changing the
         charm quark mass to 1.2 (1.8) GeV/c$^2$.
         The histograms are averages of calculations done
         at three representative $W_{\gamma p}$ values, weighted
         by the photon flux integrated over the represented
         range.
         Common systematic errors of $\cal O$(15\,\%) are not
         shown. }
\label{ds-tagged}
\end{figure}

\begin{figure}[p] \centering
\epsfig{file=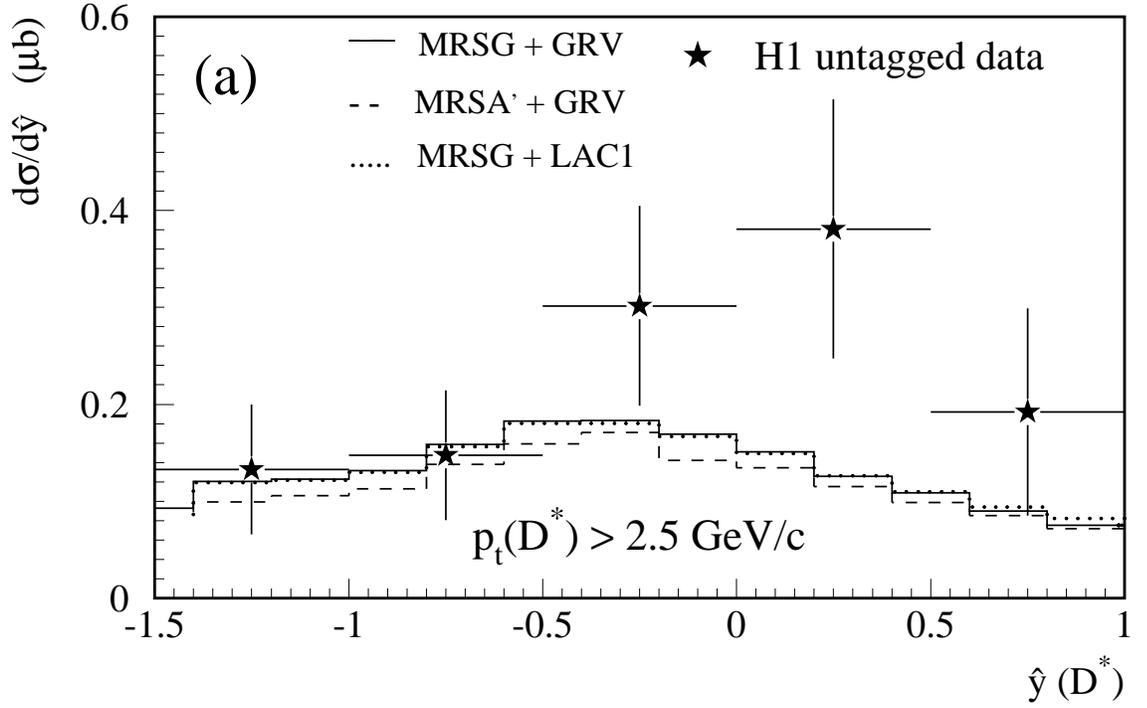,width=17.cm}
\epsfig{file=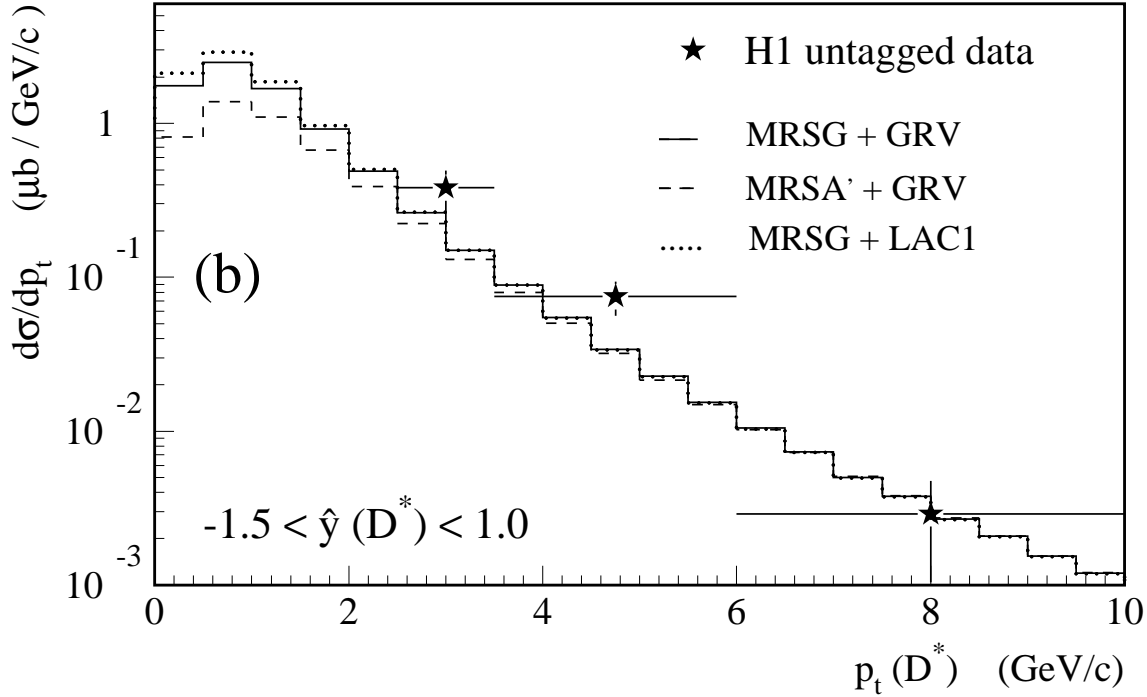,width=17.cm}
\caption[Differential eta]%
        {Differential cross sections for the untagged sample 
         (solid stars).
         (a) $1/(2 B_{c \rightarrow D^{*+}}) \cdot d\sigma 
              (\gamma p \rightarrow D^{*\pm} X)/ d\rap$
         for events with $p_t(D^*) > 2.5$ GeV/c and 
         (b) $1/(2 B_{c \rightarrow D^{*+}}) \cdot d\sigma 
               (\gamma p \rightarrow D^{*\pm} X)/ dp_t$
         for events with $-1.5 < \rap (D^*) < 1.$
         The histograms show  NLO QCD predictions for various
         parton density parametrizations for the proton and the photon:
         MRSG + GRV-G HO (solid), MRSA' + GRV-G HO  (dashed),
         and MRSG + LAC1 (dotted). A charm quark mass of
         1.5\,\gev/c$^2$ is used for the calculations.
         The histograms are averages of calculations done
         at three representative $W_{\gamma p}$ values, weighted
         by the photon flux integrated over the represented
         range.
         Common systematic errors of $\cal O$(15\,\%) are not
         shown. }
\label{ds-utagged}
\end{figure}

%
%
%
Differential photoproduction cross section distributions, 
in the visible region where no model dependent uncertainties 
enter from extrapolation,
are obtained by determining the acceptances and efficiencies
bin-by-bin, separately for the two analyses.
The distributions $1/(2 B_{c \rightarrow D^{*+}}) 
\cdot d\sigma(\gamma p \rightarrow D^{*\pm}X) /d\rap$ 
are shown in figures\,\ref{ds-tagged}a and \ref{ds-utagged}a
integrated over the range $2.5$\,GeV/c $ < p_t(D^*) < 10$\,GeV/c.
The distributions $1/(2 B_{c \rightarrow D^{*+}}) 
\cdot d\sigma(\gamma p \rightarrow D^{*\pm}X)/dp_t $ 
are presented 
in figures\,\ref{ds-tagged}b and \ref{ds-utagged}b
for the rapidity range of $-1.5 < \rap < 1.$
The results from the analyses of the tagged and untagged 
samples are not combined because they cover different 
$W_{\gamma p}$ (and $Q^2$) ranges.
Note that the largest overlap
between the two samples, namely 10\,$D^{*}$ candidates, occurs 
in the bin of $0 < \rap < 0.5$\,.
The error bars represent the statistical error 
and, for the untagged data, the bin-by-bin systematic error 
due to the trigger efficiency.  
The other systematic errors of the overall normalization
are identical to the errors quoted in table~\ref{tab-syserr}.

The histograms shown in figures\,\ref{ds-tagged} and \ref{ds-utagged} 
represent
the absolute predictions of the QCD calculation\,\cite{frixione-1}
including charm quark hadronization using the Peterson 
fragmentation function\,\cite{peterson} 
(with parameter $\epsilon = 0.06)$,
and containing both direct and resolved photon processes.
The calculations assume the parton densities MRSG for the proton
and GRV-G HO for the photon, unless stated otherwise.
The histograms are averages of calculations done
at three representative $W_{\gamma p}$ values, weighted
by the photon flux integrated over the represented range.
Good agreement within errors is observed between the shape of the 
measured $p_t$ distribution  and the NLO QCD calculation.
The $\rap$ distributions on the other hand
do not agree so well, with a possible excess of the
data in the forward direction.

To demonstrate the dependence on the charm quark mass, 
the predictions for masses $1.2\,\gev/c^2$ 
and $1.8\,\gev/c^2$ are also given in 
figure~\ref{ds-tagged} (dashed histogram).
In figure~\ref{ds-utagged} the influence of different
proton parton density functions is illustrated by overlaying
QCD predictions based on the MRSA' parametrization
(dashed histogram).
The effect of assuming the  LAC1 photon
parton density parametrization (and MRSG for the proton)
is marginal (dotted histogram). 
Although the total charm cross section 
is considerably larger when using LAC1, most of the
difference with respect to using for example 
the GRV-G HO parametrization lies in the
forward region ($2 < \rap < 4$) and at  low $p_t$,
outside of the visible range of this measurement.

\section{Diffractive photoproduction of charm quarks}%
A search for $D^{*+}$ production by diffractive processes is
performed in a sample of {\em rapidity gap} events,
in which {\it no} final state hadronic energy flow 
is detected adjacent to the proton direction. 
The selection of diffractive events 
is based on a cut in $\eta_{max} < 2$ 
and is described elsewhere\,\cite{diff}.
Here $\eta_{max}$
denotes the pseudorapidity of the most forward calorimetric 
energy deposit in excess of 400\,MeV.
The selection of $D^{*}$ candidate events 
and the rejection of contributions from DIS
are identical to those used in the
analysis of untagged data as described above.
However, because of
the small signal expected, the analysis has not been restricted 
to a particular trigger condition,
and thus comprises an integrated luminosity of 2.77\,pb$^{-1}$. 

\begin{figure}[p] \centering
\epsfig{file=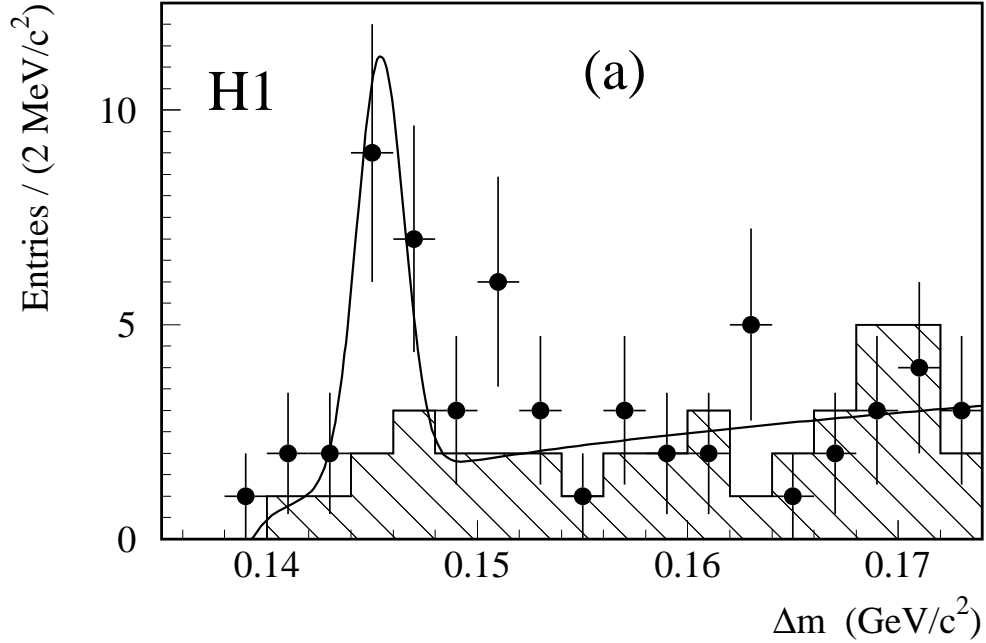,width=16cm}
\epsfig{file=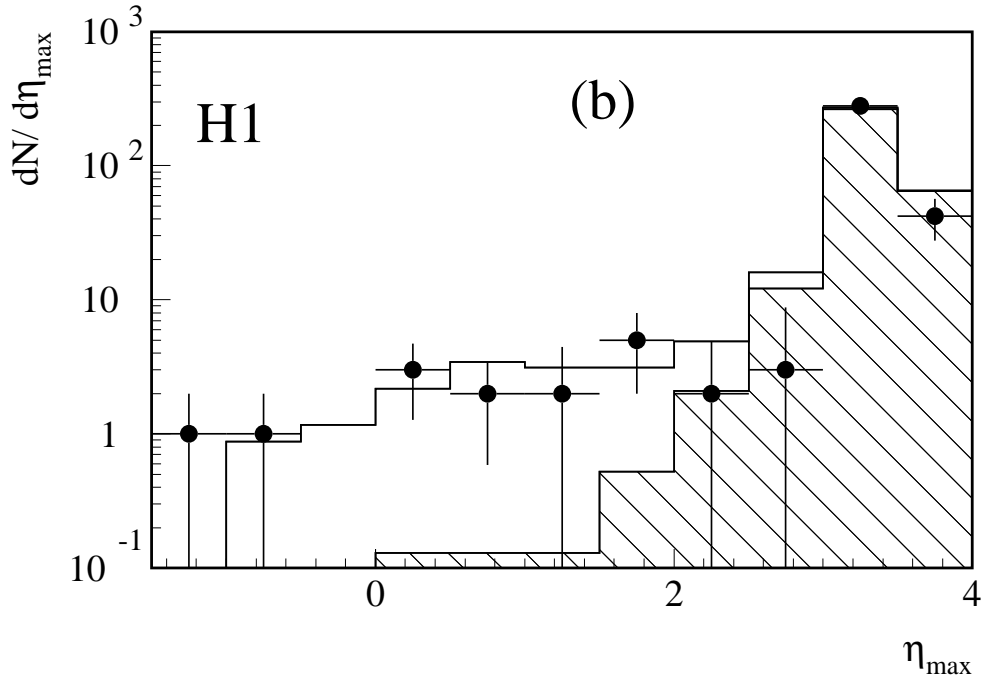,width=16cm}
\caption[Mass difference distribution]%
        {(a) Distribution  of the mass difference
         $\Delta M = M(K^-\pi^+\pi^+) - M(K^-\pi^+)$
         for events with a rapidity gap with $\eta_{max} < 2.$
         The solid dots represent the data, the hatched
         histogram indicates the background as obtained from
         wrong charge combinations. The solid line is a fit of
         a Gaussian function for the signal plus a background term. (b) 
         $\eta_{max}$ distribution of $D^{*}$ candidate events. The
         hatched histogram shows the prediction of a non-diffractive
         model (PYTHIA), the solid  histogram the 
         sum of the non-diffractive and a diffractive (RAPGAP) model.
         The former (latter) is normalized to the number of data 
         events at $\eta_{max} > 3$ ($\eta_{max} < 2$).}
\label{diff}
\end{figure}

The mass difference distribution for the selected events
is shown in figure~\ref{diff}a, exhibiting a clear $D^{*}$ signal.
To substantiate the evidence for a diffractive production
process, the $\eta_{max}$ distribution of the $D^{*}$
candidate events 
after background subtraction estimated from wrong charge
combinations, is depicted  in figure~\ref{diff}b.
The distribution shows that
most events have $\eta_{max}$ values close to the maximum 
possible value ($\eta_{max} = 3.65$), but a clear excess of events
with $\eta_{max} < 2$ is observed. The data are compared with
predictions of a non-diffractive model (as implemented in 
the PYTHIA~\cite{pythia} program)
and a hard diffractive model (RAPGAP~\cite{rapgap}).
The former (latter) is 
normalized to the number of data events at $\eta_{max} > 3$ 
($\eta_{max} < 2$). 
The sum of both models describes the shape of the data
well (solid histogram in figure~\ref{diff}b), 
while the non-diffractive model alone fails to
reproduce the shape of the $\eta_{max}$ distribution (hatched 
histogram).    

A lower limit on the visible diffractive cross section
for the kinematical region
$Q^2 < 4\,\gev^2$, $0.1 < y < 0.8$, $p_t(D^*) > 2.5$\,\gev/c,
$-1.5 < \rap < 1$ and $\eta_{max} < 2$ is derived.
Assuming the trigger efficiency in this range to be 1 
yields a conservative limit of
\begin{equation}
  \sigma(ep \rightarrow D^{*\pm} X) > 145\,{\rm pb} \hspace{1.0cm}
  {\rm at\  90 \% \  C.L.}\ \ \ \  (\eta_{max} < 2).
\end{equation}
The cross section limit can be compared with the predictions
of the diffractive model~\cite{rapgap}, which assumes a
partonic structure of the diffractive exchange. 
If the diffractive exchange is dominated by a hard gluon\,\cite{diff}
at an initial scale of $Q^2_0 \approx 4\,\gev^2$,
a cross section of 780\,pb is predicted.
On the other hand, if a quark dominated structure is assumed,
the prediction is 29\,pb in this model.
The measured cross section
is much higher then the latter prediction. 
Therefore, the data clearly disfavour a quark-dominated
diffractive exchange.

\section{Conclusions}
Charm photoproduction cross 
sections have been measured
through the detection of $D^{*\pm}$ mesons.
At an average $\gamma p$ CM energy of 200\,\gev\,
the result is  $\sigma(\gamma p \rightarrow c {\bar c} X) 
= ( 13.2 \pm 2.2 ^{+2.1}_{-1.7}\, ^{+9.9}_{-4.8})\,\mu $b
for $Q^2 < 0.01\,\gev^2$ with $\langle Q^{2}\rangle \approx 10^{-3}\, \gev^2$.
For the range $Q^2 < 4\,\gev^2$ with 
$\langle Q^{2}\rangle \approx 0.2\, \gev^2$  the values are
$\sigma(\gamma p \rightarrow c \bar{c} X)
= ( 9.3 \pm 2.1  ^{+1.9}_{-1.8}\,  ^{+6.9}_{-3.2} )\,\mu $b
at $\langle W_{\gamma p}\rangle \approx 142$\,GeV, and
$( 20.6 \pm 5.5 ^{+4.3}_{-3.9}\, ^{+15.4}_{-7.2} )\,\mu $b
at $\langle W_{\gamma p}\rangle \approx 230$\,GeV.
These values are about one order of magnitude larger
than those measured at previous fixed-target photoproduction
experiments.
Both the $W_{\gamma p}$ dependence of the photoproduction cross section 
and its dependence on  $p_t$ - and to a lesser extent on $\rap$ -
of the $D^*$ meson 
are reasonably well described by NLO QCD calculations.
A slightly better agreement is reached with a steep gluon momentum 
distribution in the proton. 
This is in accord with measurements of the 
gluon density by other methods \cite{gluon}.
The measured visible cross sections appear to be somewhat higher than the 
absolute QCD predictions.
Evidence for charm production is found in a 
subsample of events which show a distinct gap of energy flow
close to the direction of the proton and which can 
be interpreted as photon diffractive dissociation.
A quark dominated diffractive exchange is clearly disfavoured by the
present measurement.

\section*{Acknowledgments}

We are very grateful to the HERA machine group whose 
outstanding efforts have made and continue to make this
experiment possible. We thank the engineers and technicians 
for their work in constructing and now maintaining the H1
detector, the funding agencies for financial support, the
DESY technical staff for continual assistance, and the
DESY directorate for the hospitality extended to the non-DESY 
members of the collaboration.
We thank Stefano Frixione for giving advice on performing
the QCD calculations.


%

%

\end{document}